\documentclass[prl,twocolumn,aps,superscriptaddress,showpacs,preprintnumbers]{revtex4}
\usepackage{amsfonts}

\usepackage{varioref,exscale,latexsym,amsmath,amssymb,wasysym,soul}

\usepackage{graphicx}
\usepackage{bm}
\usepackage{slashed}
\usepackage[colorlinks=true, pdfstartview=FitV, linkcolor=black, citecolor=black, urlcolor=black]{hyperref}

\newcommand{\mockbarsaturn}{\saturn}

\newcommand{\mockbar}{\mockbarsaturn}

\begin{document}

\newcommand{\be}{\begin{equation}}
\newcommand{\ee}{\end{equation}}
\newcommand{\beq}{\begin{eqnarray}}
\newcommand{\eeq}{\end{eqnarray}}
\newcommand{\bea}{\begin{eqnarray}}
\newcommand{\eea}{\end{eqnarray}}
\newcommand{\beqn}{\begin{eqnarray}}
\newcommand{\eeqn}{\end{eqnarray}}
\newcommand{\rd}{\mathrm{d}}
\newcommand{\X}{\mathbb{X}}
\newcommand{\A}{\mathbb{A}}
\renewcommand{\P}{\mathbb{P}}
\newcommand{\ack}[1]{{\bf Pfft. #1}}
\newcommand{\pa}{\partial}
\newcommand{\osigma}{\overline{\sigma}}
\newcommand{\orho}{\overline{\rho}}
\newcommand{\myfig}[3]{
\begin{figure}[ht]
\centering
\includegraphics[width=#2cm]{#1}\caption{#3}\label{fig:#1}
\end{figure}
}
\newcommand{\littlefig}[2]{
	\includegraphics[width=#2cm]{#1}}
\newcommand{\1}{{\rm 1\hspace*{-0.4ex}%
\rule{0.1ex}{1.52ex}\hspace*{0.2ex}}}

\def\pa{\partial}
\def\dd{\.\cdot \.}
\def\S{\mathbb{S}}
\def\pp{\cal{P}}

\def\N{\nabla}
\def\pa{\partial}
\def\om{\omega}
\def\dd{\!\cdot \!}
\newcommand{\un}[1]{\underline{#1}}
\newcommand{\ackm}[1]{\marginnote{\small Pfft!: #1}}

\def\dd{\!\cdot \!}
\def\S{\mathbb{S}}
\def\s{\sigma}
\def\he{\hat{\epsilon}}
\def\P{\cal{P}}
\def\Q{\mathbb{Q}}
\def\bra{\langle}
\def\ket{\rangle}
\def\tr{\mathrm{tr}}
\def\nn{\nonumber}
\def\bP{\bm{P}}
\def\PP{\mathbb{P}}

\title{Emergent ``Quantum'' Theory in Complex Adaptive Systems}


\author{Djordje Minic}
\affiliation{Department of Physics, Virginia Tech, Blacksburg, VA 24061, U.S.A.}

\author{Sinisa Pajevic}
\affiliation{Mathematical and Statistical Computing Laboratory,
Center for Information Technology,
National Institutes of Health, 
Bethesda, MD 20892-5620, U.S.A. }

\date{\today}

\begin{abstract}
Motivated by the question of stability, in this letter we argue that an effective quantum-like theory can emerge in
complex adaptive systems.  In the concrete example of stochastic Lotka-Volterra dynamics, the relevant effective
``Planck constant'' associated with such emergent ``quantum'' theory has the dimensions of the square of the unit of
time. Such
an emergent quantum-like
theory has inherently non-classical stability
as well as coherent properties that are not, in principle, endangered by thermal fluctuations
and therefore might be of crucial importance in complex adaptive systems.
\end{abstract}


\maketitle

{\it Introduction:} That many-body systems can harbor emergent phenomena not present in the microscopic description
of such systems has been well known in the literature on condensed matter physics \cite{wen}. For example, emergent
quantum critical phenomena at low temperatures can be described by effective quantum field theories, which
are not inherent in the microscopic description of actual materials that exhibit such emergent phenomena \cite{wen}.
In this letter we argue that effective ``quantum'' theory can emerge in complex adaptive systems, even though the
microscopic dynamical description of such systems is stochastic and dissipative.  We emphasize that in our case we
are talking about an emergent ``quantum'' framework (see also \cite{tony, droplet, fluidpilot, eqbook}), which can be used to define different effective theories.
The non-reductionist nature of emergent phenomena has been already emphasized in the past \cite{anderson}.  Usually
one invokes energetic arguments for the emergence of certain correlated states of matter.  In our proposal the
crucial distinguishing aspect of emergent ``quantum'' theory would be the emergent ``quantum'' stability, as implied
by the emergent ``quantum'' coherence. We emphasize the unique nature of quantum stability and distinguish it from
its classical counterpart (whether in the classically deterministic or stochastic contexts).  Also, as shown in the
context of quantum information and quantum computation, the informational and computational consequences of quantum
coherence cannot be modeled by purely classical (either deterministic or stochastic) systems \cite{qcqi}.  Thus the
computational aspects of emergent ``quantum'' systems would be quite unique, as opposed to their classical
counterparts.  Therefore, it might be advantageous for complex adaptive systems on various grounds (stability,
informational and computational aspects) to display emergent ``quantum'' behavior in certain regimes (defined by an
emergent, or ``mock'', Planck constant $\mockbar$.)  Our proposal regarding the emergence of effective ``quantum''
theory in complex adaptive systems should be understood as a working hypothesis, worthy of further exploration and
elaboration.  In what follows we discuss various arguments for such an emergent mock quantum theory and argue that
the new non-classical kind of stability associated with such an emergent quantum behavior should be natural and
advantageous for adaptable complex systems \cite{bohr}. After presenting our main argument for the emergence of
effective ``quantum'' theory in complex adaptive systems, we also point out that such an emergent mock quantum
theory is natural in non-linear complex dissipative systems, as pointed out by 't Hooft in a completely different
physical context of quantum theory of gravity \cite{thooft}. We adapt 't Hooft's reasoning in order to argue, once
again, for the emergence of effective ``quantum'' theory in complex adaptive systems. Finally, we illustrate our
main point about the emergence of mock ``quantum'' theory in the context of the stochastic Lotka-Volterra dynamics.
We emphasize that our proposal can be seen as a very natural extension of the recent experimental and theoretical work on
the emergent quantum-like physics found in the context of fluid dynamics \cite{droplet, fluidpilot}.

{\it Emergent ``Quantum'' Theory:} We start addressing the technical content of our proposal by representing a given
deterministic model  (e.g. the Lotka-Volterra system) in the Hamilton-Jacobi picture, 
$
\frac{\partial S}{\partial t} +H(Q_i,P_i \equiv \frac{ \partial S}{\partial Q_i})=0. 
$
Here, the action is defined by the canonical expression ($Q_i$ denotes the configuration variables and $P_i$ their conjugate momenta)
$
S= \int (P_i \dot{Q}_i - H) dt.
$
For such a general deterministic model described in the Hamilton-Jacobi picture, the phase space volume is conserved (the volume density of phase space being $\rho$)
$
\frac{\partial \rho}{\partial t} + \nabla(\rho \vec{v}) =0,
$
where the velocity $\vec{v}$ is defined as $v_i \equiv \dot{Q}_i$, and in the single particle case $\vec{v} = \frac{\nabla S}{m}\equiv \frac{\vec{P}}{m}$.
Inspired by the formalism of Rosen and Schiller \cite{rs} we define the following new variables
\be
\psi \equiv \pm \sqrt{\rho} \exp(i \frac{S}{\mockbar}).
\label{psi}
\ee 
The ``mock'' (or effective) Planck constant $\mockbar$ appears by dimensional arguments given this definition and
the dimensionality of the action $S$.  For example, in the case of the Lotka-Volterra dynamics the dimension of the
action is $\tau^2$, $\tau$ having the dimension of time. In the next section we apply our mock-quantum formalism to
the specific example of the Lotka-Volterra model \cite{lv, other}, but for now we keep our discussion more
general. The variable $\psi$ in Eq. \ref{psi} satisfies the following ``$\psi$-equation'' \cite{rs} 
\be i \mockbar
\frac{\partial \psi}{\partial t} = H (Q_i, -i \mockbar \frac{\partial}{\partial Q_i}) \psi + V_Q (\psi, \psi^*)
\psi.
\label{mq1}
\ee There is also the complex conjugate equation for $\psi^*$.  Note that in this expression $-V_Q$ is the so-called
``quantum'' potential (of de-Broglie and Bohm \cite{holland, bohm, rpf}), which depends on $\psi$ and $\psi^*$ and
comes from the kinetic term in the action. In the case of the canonical (single particle) kinetic term
$\frac{P_i^2}{2m}$, the quantum potential is $V_Q = \frac{{\mockbar}^2}{2m} \frac{\nabla^2
  (\sqrt{\rho})}{\sqrt{\rho}}$.  As pointed out in a different context, in which the real Planck constant $\hbar$
appears instead of its ``mock'' counterpart $\mockbar$ \cite{rs}, the above equation (\ref{mq1}) does look like a
non-local and non-linear ``Schr\"{o}dinger-like'' equation.  (Obviously the canonical Schr\"{o}dinger equation is
given by setting $V_Q\equiv 0$.)  This non-local and non-linear dynamics evolves, in general, pure states into mixed
states, and thus, it is not coherent (and not unitary) in the sense of the canonical linear ``quantum'' theory \cite{rs}.

This procedure of rewriting the usual classical equations of motion in terms of
a non-linear (and non-local) generalization of the usual Schrodinger equation goes back to
the work of Rosen and Schiller. The essential idea is easy to understand
once one recalls the de-Broglie-Bohm rewriting of the canonical Schrodinger equation in terms of
the action variable (the phase of the wave function) and the probability density (the square of the
modulus of the wave-function). In that familiar case one gets the usual classical equations of motion
in which the canonical potential is corrected by the quantum potential. What Rosen and Schiller observed is that this procedure can be inverted and that the classical equations of motion can
be rewritten as a Schrodinger equation in which a new potential (i.e. the quantum potential)
is added to the Hamiltonian featuring in the Schrodinger equation, making the resulting
Schrodinger-like equation non-linear and non-local (because of the explicit dependence of the quantum potential on the wave-function).

The above rewriting is true for deterministic systems (such as the Lotka-Volterra dynamics), and in such systems the
dynamics can be dominated by a single attractor or a global minimum, yielding systems that lack flexibility. In a
realistic complex system such deterministic dynamics should be supplemented with a source of noise, or other
environmental factors not included in the model, ${\eta}$, in which case the stability analysis is generalized from
the deterministic to stochastic analysis \cite{stability}. The presence of multiple minima and the stochastic
transitions brings the concepts of metastability and multi-stability that enrich the dynamics of complex systems
\cite{kelso2012}. Nevertheless, in such classical forms of stability there is inherent dichotomy between the
stability and control on one side and the flexibility on the other. The control of multistable systems is of great
practical importance and an active area of research \cite{pisarchik}.

In this work we propose a novel (``quantum'') type of stability in which the ``quantum'' constraints and linearity can
provide stability and flexibility at the same time, and we argue that this new type of stability can emerge in complex systems if the system
and/or the environmental factors are adaptive. Biological systems provide perfect examples of such adaptive
behavior. In a cell, any metabolic process or a pathway is strongly coupled to the homeostatically regulated
intracellular environment. Ever since the enclosure of self-replicating RNA in a membrane, the pathways and the
intracellular environment are evolving simultaneously and their interaction, while extremely important, is not
understood to this day. We argue that it is advantageous for an adaptive complex system to develop this new type of
``quantum'' stability and linearity and show how it can emerge if the environmental/stochastic source $\eta$ cancels
the non-linear and non-local part of the ``$\psi$-equation'', turning it into an emergent and effective
Schr\"{o}dinger equation. In the presence of the environmental/noise term $\eta$, the equation (\ref{mq1}) should be modified to
\be i \mockbar \frac{\partial \psi}{\partial
  t} = H (Q_i, -i \mockbar \frac{\partial}{\partial Q_i}) \psi + \underbrace{V_Q (\psi, \psi^*) \psi +
 {\eta}}_{\approx 0},
\label{cancel}
\ee where we have indicated that the $V_Q (\psi, \psi^*) \psi$ and ${\eta}$ can be combined into one effective term
which according to our proposal cancels in a complex adaptive environment. Note that such cancellation would be
difficult to achieve if $\eta$ is purely stochastic and hence the environment too is expected to be adaptive. Since
$\eta$ is expected to contain a stochastic component we discuss later the consequences of imperfect
cancellation in Eq. \ref{cancel}. In our proposal, the environmental factors crucially
depend on the ``holistic'' description of the system, and hence a complex adaptive system can adjust ``holistically''
to the adaptive environment (and vice versa), leading to the emergence of an effective ``quantum''
dynamics. However, given the underlying classical origin of our ``mock'' quantum, the Bell inequality is  not violated. 
 (Note that this is somewhat reminiscent of
the observed de-Broglie-Bohm-like behavior of guided droplet ``particles'' in a vibrating bath of classical fluid \cite{droplet, fluidpilot}.
In particular, the above cancelation mechanism is directly analogous to the ``harmony of phases'' \cite{fluidpilot} experimentally found in the emergent quantum-like physics in the context of fluid mechanics. Such
systems also do not violate the Bell inequality.)
The process by which
mock-quantum framework emerges is in essence the reverse of the decoherence approach to the quantum-to-classical transition \cite{zurek},
as the system as a whole ``re-coheres''. Obviously, this could happen only if $\eta$ depends on $\psi$, or, in other
words, if the environment is adaptive and ``holistic'' (i.e. depends on $\psi$ and $\psi^*$). In this case, the
holistic ``mock quantum'' wave function $\psi$, by its definition (\ref{psi}), can be adjusted to the ``holistic''
changes in an adaptive environment represented by the noise term $\eta$, and vice versa, thus ensuring the
robustness of the emergent ``quantum'' description.  The more fundamental reason for the advantageous nature of
emergent ``quantum'' theory might be found in the linear structure of quantum theory, which is tightly related to
the concept of of maximally symmetric Fisher information \cite{wootters,chia}.  The adaptive dynamics of complex
systems would thus adjust the whole system so that only the linear part of the above ``$\psi$-equation'' form of the
complex dynamics is left. This selection mechanism would then lead to effective or ``mock'' quantum theory 
\be i
\mockbar \frac{\partial \psi}{\partial t} = H (Q_i, -i \mockbar \frac{\partial}{\partial Q_i}) \psi.  
\ee
Note that, given what we know about canonical quantum theory, here $\psi$ represents an emergent {\it amplitude of probability}.
This is radically different of the well-known rewritings of stochastic equations in terms of a Schr\"{o}dinger-like equation for classical
probability. We emphasize that this emergent quantum theory is distinct from the underlying quantum theoretic description at the molecular biochemical level, given the usual decoherence effects \cite{zurek}.

We now address the question of dissipation in complex adaptive systems. The issue of emergence of quantum theory in classical dissipative systems has been (perhaps surprisingly) discussed in a completely different context
of the fundamental physics of quantum gravity \cite{flm}, by 't Hooft \cite{thooft}.
Note that there exist other discussions of emergent quantum theory  \cite{othereq}. However,  't Hooft's discussion of emergent quantum theory is particularly interesting for us because of its crucial emphasis on dissipation. 
Translated into our context, 't Hooft's argument runs as follows.
The ``mock'' Schr\"{o}dinger equation:
$
\frac{\partial \psi}{\partial t} = - i\Omega\psi, \quad \Omega \equiv \frac{H}{\mockbar},
$
can be reproduced by 't Hooft's system:
$
\frac{d \varphi(t)}{d t} = \omega(t), \quad \frac{d \omega(t)}{d t} = - \frac{\kappa}{2} \frac{d f^2}{d \omega}, \quad f(\omega) \equiv det(\Omega - \omega),
$
where $\varphi$ is valued between $0$ and $2\pi$ and $\psi$ is periodic in $\varphi$:
$\psi(\omega, \varphi) = \psi(\omega, \varphi + 2\pi)$, and the function
$f(\omega)$ has zeros at the eigenvalues of $\Omega$.
't Hooft's point is that the eigenvalues of $\Omega$ get attracted to the zeros $\omega_i$
exponentially in time, as the system enters its limit cycle \cite{thooft}.
Note that 't Hooft
is concerned with the emergence of the canonical quantum theory from some more fundamental framework (such as quantum gravity) and he
does not discuss emergent quantum theory in complex adaptive systems, which is our main concern.
However, given what we have already said in this section, we see that 't Hooft's formalism can serve as a precise illustration of our central idea regarding the
emergence of ``mock'' quantum theory in dissipative and adaptive complex systems.
We can use this proposal in our context to argue that the deterministic complex adaptive system,
characterized by an adaptive environment, described by the above dissipative system of 't Hooft, can
lead to an emergent mock-quantum theory. To argue this we use the action-angle canonical variables:
$
I \equiv \int P_i dQ^i, \varphi \equiv \frac{\partial S}{\partial I}.
$
In the absence of 't Hooft's dissipation $\frac{d \omega(t)}{d t} =0$, we have the canonical equations
$
\frac{d I}{d t} \equiv  - \frac{\partial H}{\partial \varphi} =0, $ and $
\frac{d \varphi(t)}{d t} \equiv  \frac{\partial H}{\partial I} = const.
$
Thus, in the presence of an environment modeled by the above 't Hooft's dissipation function $f^2$ one
can, following 't Hooft's argument \cite{thooft}, expect the emergence of effective ``quantum'' theory.
Once again, this example indicates that our proposal for emergent ``quantum'' theory in complex adaptive systems is in principle realizable. 
Note that, more recently, 't Hooft has argued for a cellular automaton interpretation of canonical quantum theory \cite{thooft1}.
Given what is known about the importance of cellular automata for modeling of complex adaptive systems \cite{cellular}, it is natural to apply this most recent 't Hooft's reasoning \cite{thooft1} to generic complex adaptive systems and, once again,
argue for the emergence of  ``quantum'' theory in which the effective ``Planck constant'' is given by dimensions of the dynamical action that describes the particular complex system.
Finally, let us address briefly the question of robustness of the emergent ``quantum'' theory in complex adaptive systems. In particular, one might
ask what happens to the emergent Schr\"{o}dinger equation under imperfect cancellation in 
Eq. \ref{cancel}, i.e., by adding a perturbation to the ``mock'' Schr\"{o}dinger equation.
The natural proposal here is that precisely such perturbations will lead to the ``collapse''
of the emergent ``wave function'' and the actually observed values of measured quantities
(with the probability distribution governed by the Born rule). Thus the perturbations in (adaptive) noise would
be crucial for a dynamical ``collapse'' of the emergent Schr\"{o}dinger ``wave function'' along the lines
of various proposals reviewed in \cite{bassi}. (The crucial role of noise in the emergence of classical behavior in the
de-Broglie-Bohm interpretation of the canonical quantum theory is nicely summarized in \cite{bohm}.)

A related question is if the mock-quantum framework described here can be applied to the non-Hamiltonian systems. In the case of non-Hamiltonian systems, one would need to separate the Hamiltonian piece from the rest of dynamical equations in question (such as the dissipative piece). The non-Hamiltonian piece, as well as the imperfect cancellation will then act as pertrurbations of the ``mock quantum'' dynamics, which in turn will lead to the ``collapse'' of the emergent ``wave function'' and the actually observed values of measured quantities (with the probability distribution governed by the Born rule). 
Alternatively, the dissipative piece can be significantly reduced through cancelation in a similar way in which the dynamical non-linearity cancels the dissipative and dispersive effects in the formation of the dissipation solitons. The emergence of mock-quantum, however, will only be possible for a narrow range and type of environmental conditions. Below, we illustrate our proposal in the example of the Lotka-Volterra system.

{\it The Lotka-Volterra example:}
Now we illustrate our proposal by considering the Lotka-Volterra system.
The deterministic Lotka-Volterra system \cite{lv} specifies the number of species and how their populations interact
and change in time. 
For any number of speciesl with population $N_i (t)$ the Lotka-Volterra equations read \cite{lv, other}
\be
\dot{N_i} = \epsilon_i N_i  - \beta_i^{-1} \sum_{j=1}^{m}\alpha_{ij} N_i N_j ,
\ee
where $\dot{N_i}$ denotes the time ($t$) derivative, $\epsilon_i$ is the
relevant autoincrease or autodecrease parameter, $\beta_i$ is Volterra's equivalent number parameter and $\alpha_{ij}$
($\alpha_{ij} >0 $ and $\alpha_{ij} = -\alpha_{ji}$)
denotes the interaction strength between the $i$ and the $j$ species \cite{lv, other}.
The stationary ($\dot{N_i}=0$)  population levels $N_i = q_i$ occur for $\epsilon_i \beta_i +\sum_{j=1}^{m}\alpha_{ij} q_j=0$, where
it is usually assumed that $\alpha_{ij}$ is nonsingular and $q_i >0$ \cite{other}.
The classical Hamiltonian for the Lotka-Volterra can be
written as follows \cite{other}. First we introduce
$
z_i \equiv \log( N_i/q_i) .
$
Then the above deterministic Lotka-Volterra equations read as follows \cite{other}
\be
\dot{z_i}= \sum_{j=1}^{m}\gamma_{ij} \alpha_j (e^{z_j} -1), \quad \gamma_{ij} = -\gamma_{ji},
\ee
where $\gamma_{ij} \equiv \alpha_{ij} \beta_i^{-1} \beta_j^{-1}$ and $\alpha_i \equiv q_i \beta_i$.
For the two-species case the canonical variables are
$
z_1= Q, \quad z_2=P,
$
and the Lotka-Volterra Hamiltonian is \cite{other} 
\be
H = \gamma \alpha_1 (e^Q - Q) + \gamma \alpha_2 (e^P -P).
\label{hlv}
\ee
This deterministic Lotka-Voterra equations can be rewritten as the canonical Hamilton equations:
$
\dot{Q} = \frac{\partial H}{\partial P}, \quad \dot{P}= -\frac{\partial H}{\partial Q},
$
with $\gamma_{12}=-\gamma_{21} \equiv \gamma$, or in terms of the Hamilton-Jacobi equations, or simply in terms of $N_i$ variables. 

Now we discuss the emergent mock-quantum theory in this specific case. According to our proposal, one first rewrites the deterministic
Lotka-Volterra system in the Hamilton-Jacobi picture by using the variables $\psi$ and $\psi^*$. Then one considers
the stochastic Lotka-Volterra system in an adaptive environment in order to argue for the emergence of a
quantum-like, linear, description in terms of $\psi$ and $\psi^*$.  The emergent ``mock quantum'' Lotka-Volterra
dynamics is endowed with the properties we expect from a quantum-like theory.  For example, in the emergent quantum
Lotka-Volterra system there exists the ``uncertainty'' principle which asserts that the coordinates (integrals of
the numbers of species) and their conjugate momenta cannot be simultaneously determined. This new fundamental
feature should provide a striking experimental signature of ``mock'' quantum theory in the context of the
Lotka-Volterra dynamics.  Similarly one would be only able to talk about transition probabilities from a certain
number of species to another number of species. One would also have the zero point energy, tunneling and interference as
the hallmarks of a quantum-like theory.  In the stationary case, the emergent quantum theory
of Lotka-Volterra dynamics advocated in this letter is characterized by the emergent stationary Schr\"{o}dinger
equation $ H \psi_n = E_n \psi_n, $ where $E_n$ are the emergent eigenvalues of $H$ and where as usual, $\psi_n(t) =
\exp(-\frac{i}{\mockbar} E_n t)\psi_n(0)$, $n=0,1,2...$.  Now, for small $P$ and $Q$, which is the linearized approximation of the
slow Lotka-Volterra dynamics, $e^Q \sim 1+ Q +Q^2/2$ and similarly $e^P \sim 1+ P +P^2/2$ , and the effective
Lotka-Volterra dynamics is harmonic \cite{other} \be H \sim \gamma (\alpha_1+\alpha_2) + \frac{1}{2} (\gamma
\alpha_1 Q^2 + \gamma \alpha_2 P^2).  \ee The ``mock'' quantum theory is then mapped to the simple quantum-like
harmonic oscillator with energy levels given by the standard result for the ``energy spectrum'' 
\be E_n = \gamma
(\alpha_1+\alpha_2) + \mockbar \gamma \sqrt{\alpha_1 \alpha_2} (n +\frac{1}{2}), 
\ee 
where $\mockbar \equiv \tau^2$,
the dimension of the Lotka-Volterra action.  The corresponding ``stationary wavefunctions'' $\psi_n$ are given in
terms of the appropriate Hermite polynomials.  In this case all transition probabilities can be computed exactly.
Similarly the classical variables $Q$ and $P$ become the canonically normalized operators $\hat{Q}$ and $\hat{P}$,
obeying the canonical commutation relations $[\hat{Q}, \hat{P}] = i \mockbar$. For a given ``quantum'' state, this commutation
relation in turn implies the canonical uncertainty relation between $\Delta Q$ and $\Delta P$. Therefore in the
emergent ``quantum'' phase the Lotka-Volterra variables $Q$ and $P$ and therefore $z_1$ and $z_2$, cannot be
simultaneously measured.
In the case of the above harmonic oscillator (H.O) approximation, for the case of
the stationary states with the energy $E_n$, the quantum potential is simply given by \cite{holland}
\be
V_{Q}^{(H.O.)} = \mockbar \gamma \sqrt{\alpha_1 \alpha_2} (n +\frac{1}{2}) - \frac{1}{2} \gamma   \sqrt{\alpha_1 \alpha_2}Q^2.
\ee
This follows from the exact solution of the ``mock Schr\"{o}dinger equation'', which
involves the standard Hermite polynomials.
The time dependent form of the ``mock-quantum potential'' in our situation can be also easily inferred 
from the explicit solution of the canonical quantum harmonic oscillator \cite{holland}
\be
V_{Q}^{(H.O.)} (t) = -\frac{\gamma^2}{2}  \alpha_1 \alpha_2 (Q - A \cos{(\gamma   \sqrt{\alpha_1 \alpha_2} t)} )^2 +
\frac{\mockbar}{2} \gamma \sqrt{\alpha_1 \alpha_2}.
\label{vqharm}
\ee
This expression represents the time-dependent ``holistic'' response of the environment in
our proposal, i.e. it determines the negative of the ``holistic noise'' term $\eta$.
(Recall that $\psi V_{Q}^{(H.O.)} (t) + \eta \to 0$, is a necessary criterion for the emergence of effective mock quantum theory,
in this case, in the harmonic oscillator approximation of the stochastic Lotka-Volterra dynamics.)

We remark that the effective (mock) Planck constant is introduced by dimensional arguments in the
rewriting of the classical dynamics in terms of an effective wave function. Because of the fact that the action for that classical dynamics features in the phase of the effective wave function, something has to compensate the dimensionality of that action in the dimensionless phase. This consideration is quite general and it does not rely on a having a simple harmonic oscillator-like system as an actual example. The physical meaning of such an effective Planck constant is tied to the physical meaning of the action, and the actual value of the mock Planck constant is system-dependent and not in general, universal (unlike the fundamental Planck constant from canonical quantum theory).

{\it Note that the crucial prediction of the emergent quantum theory in the Lotka-Volterra system in an adaptive
  environment is the emergent quantum stability of the discrete energy levels as well as the existence of the
  associated vacuum energy.} The transitions between discrete energy levels could be induced by coupling the
Lotka-Volterra system to another emergent quantum system. Similarly, the Lotka-Volterra oscillations, which in our
case represent cycles in a phase space, get quantized. In general, beyond the harmonic oscillator approximation, we
have to solve the mock quantum system (given by the above $H$ in (\ref{hlv})) numerically.  (One could also consider
the small $P$ limit for arbitrary $Q$, in which case a WKB analysis is possible.)  The harmonic approximation is
mainly interesting from the conceptual point of view, because it illustrates the emergence of discrete energy levels
and thus the associated ``quantum'' stability that we propose. However, it might be also interesting from a more
practical point of view, since our proposal could be tested on a complex system operating in the harmonic oscillator
limit of the Lotka-Volterra system.  We see that the environmental ``holistic noise'' in this simplified harmonic
oscillator case is controlled by a simple oscillatory (harmonic) function that depends on variable $Q$ and the wave
function $\psi$, which for the ground state is a simple Gaussian. This means that in the case of two interacting
species, whose concentrations are dually related, as the momentum is to position in the case of harmonic oscillator,
an appropriate oscillatory environment can be obtained that would lead to the emergence of the mock quantum harmonic
oscillator (with stable ``energy'' levels). We note that oscillatory phenomena are ubiquitous in biology, both at
tissue and at cellular level \cite{cellosc}, and are essential for the functioning of biological systems. It is then
tempting to conjecture that by investigating such oscillations, a potential evidence for the necessity and emergence
of mock quantum stability can be demonstrated. While real biological systems might be too complicated to analyze, a
simplified synthetic regulatory cell or an auto-catalytic chemical process, in which the nature of these
oscillations can be controlled, could be a platform for investigating the relevance of our proposal \cite{synthetic}.
Essentially what we propose as a test of our theoretical discussion should be seen as a direct analog of the recent quantum-like fluid dynamics experiments \cite{droplet, fluidpilot}, however, to be conducted in the context of synthetic biology.

Our claim would be that there should exist an emergent time scale $\tau$ at which
the system would have to be described by an emergent ``quantum'' dynamics proposed in this letter. 
In particular, the emergent ``quantum'' dynamics would imply a new non-classical stability in the Lotka-Volterra
system, which cannot be modeled by the usual methods of non-linear and stochastic dynamics \cite{stability}.
Finally, we note that the formal quantization of the Lotka-Volterra system (using the usual Planck constant $\hbar$)
has been discussed previously in the literature \cite{enz}. However, that discussion does not have anything to do
with the central point of this letter regarding the emergence of effective quantum theory in complex adaptive
systems.  Next, we compare our proposal to the approach to quantization known as stochastic quantization
\cite{parisi}. In this approach (Euclidean) quantization is viewed as a stationary (equilibrium) limit of a
fictitious stochastic process (with an extra fictitious evolution parameter) whose effective ``temperature'' is
given by the Planck constant. (Even in our case the Euclidean version of our effective, mock quantum theory, seen
from the viewpoint of its Euclidean Feynman path integral, can be also obtained as a stationary limit of a
fictitious stochastic process, characterized by an extra fictitious evolution parameter, with the crucial difference
that in our case the effective Planck constant is emergent and given by $\mockbar$.)  Note that in our discussion of
effective, mock, quantum theory, the adaptive environmental part is crucially needed to cancel the non-linear and
non-local part of the classical dynamics rewritten in terms of somewhat unusual ``wave-function''-like variables, thus
leading to emergent ``quantum'' dynamics in which the evolution parameter corresponds to the real, physical time.

{\it Conclusion:} We offer some general comments that should be useful for further investigations of our proposal:
1) One natural interpretation for the proposed ``mock''' quantum theory could be understood as a reverse analog of
``quantum Darwinism'' of Zurek \cite{zurek} considered in the context of decoherence models of the canonical
quantum-to-classical transition, since in our proposal the effective quantum behavior emerges.  2) The dimensions
and the value of the fundamental parameter, $\mockbar$, are system dependent (e.g., different values of $\tau$ for
different Lotka-Volterra systems).  3) In many situations one considers discrete Lotka-Volterra system of equations
as opposed to the continuous one; it is natural to ask whether the parameter $\tau$ can be constrained from the
point of view of discrete Lotka-Volterra dynamics.  4) Finally, the proposed emergent ``mock'' quantum theory opens
the door for various collective many-body ``mock'' quantum effects in biological systems.  Our proposal can be
intuitively understood as a new type of stability in biological systems, but should be clearly distinguished from
the arguments that canonical quantum physics is relevant in biological systems \cite{quantumbio}. Our effective
``mock quantum'' theory comes with a new fundamental deformation parameter (e.g., the time scale in the
Lotka-Volterra models) that is emergent and thus distinguishable from the canonical fundamental quantum theory which
usually suffers in competition with realistic thermal biological environments. Obviously in this note we have only
scratched the surface \cite{supermq} and further work is needed to understand the full implications of our proposal

{\it Acknowledgments:}
We thank V. Balasubramanian, N. Goldenfeld, W. Mather, I. Nemenman, M. Pleimling, V. Scarola, U. Tauber and J. Xing for illuminating discussions and comments.
D. M. thanks the Aspen Center for Physics for providing a stimulating environment for work.
DM is supported in part by the 
US Department of Energy
under contract 
DE-FG02-13ER41917, task A.


\end{document}